\documentclass[12pt,showpacs,preprintnumbers,amsmath,amssymb,floatfix]{revtex4}
\usepackage[pdftex]{color}
\usepackage{epsfig} 
\usepackage {array} 
\usepackage {amsmath} 
\usepackage {lscape}
\usepackage{graphicx}
\begin{document}
\preprint{}
\title{Parametric instability of the helical dynamo}
\author{Marine Peyrot$^{1,2}$, Franck Plunian$^{1,2}$, Christiane Normand$^3$}
\email{Marine.Peyrot@ujf-grenoble.fr, Franck.Plunian@ujf-grenoble.fr, Christiane.Normand@cea.fr}
\affiliation{$^{1}$ Laboratoire de G\'eophysique Interne et Tectonophysique, CNRS, Universit\'e Joseph Fourier,
Maison des G\'eosciences, B.P. 53, 38041 Grenoble Cedex 9, France \\
$^{2}$ Laboratoire des Ecoulements G\'{e}ophysiques et
Industriels, CNRS, Universit\'e Joseph Fourier, INPG, B.P. 53, 38041 Grenoble Cedex 9, France \\
$^3$Service de Physique Th\'eorique, CEA/DSM/SPhT, 
CNRS/URA 2306, CEA/Saclay, 
91191 Gif-sur-Yvette Cedex, France}
\date{\today}
\begin{abstract}
We study the dynamo threshold of a helical flow
made of a mean (stationary) plus a fluctuating part.
Two flow geometries are studied, either (i) solid body or (ii) smooth. 
Two well-known resonant dynamo conditions, elaborated for stationary helical flows
in the limit of large magnetic Reynolds numbers,
are tested against lower magnetic Reynolds numbers and for fluctuating flows (zero mean).
For a flow made of a mean plus a fluctuating part the dynamo threshold 
depends on the frequency and the strength of the fluctuation.
The resonant dynamo conditions applied on the fluctuating (resp. mean) part seems to be a good diagnostic
to predict the existence of a dynamo threshold when the fluctuation level is high (resp. low).
\end{abstract}
\pacs{47.65.+a}
\maketitle

\section{Introduction}
In the context of recent dynamo experiments \cite{Cardin02,Bourgoin02,Frick02},
an important question is to identify the relevant physical parameters which control the dynamo threshold and 
eventually minimize it. In addition to the parameters usually considered, like the geometry of the mean flow
\cite{Ravelet05,Marie06} or the magnetic boundary conditions \cite{Avalos03, Avalos05}, 
the turbulent fluctuations of the flow seem to have an important influence on the dynamo threshold \cite{Leprovost05,Fauve03,Laval06,Petrelis06}. Some recent experimental results \cite{Ravelet05b,Volk06} suggest that the large spatial scales of these fluctuations could play a decisive role.

In this paper we consider a flow of large spatial scale, fluctuating periodically in time, such that its geometry at some given time is helical. Such helical flows have been identified to produce dynamo action \cite{Lorz68,Ponomarenko73}. Their efficiency has been studied in the context of fast dynamo theory
\cite{Roberts87,Gilbert88, Ruzmaikin88, Basu97, Gilbert00, Gilbert03} and they have led to the realization of several dynamo experiments \cite{Gailitis80, Gailitis00, Gailitis01, Frick02}.

The dynamo mechanism of a helical dynamo is of stretch-diffuse type.
The radial component $B_r$ of the magnetic field is stretched to produce a helical field (0, $B_{\theta}, B_z$),
where ($r, \theta, z$) are the cylindrical coordinates. The magnetic diffusion of the azimuthal component $B_{\theta}$ produces some radial component $B_r$ due to the cylindrical geometry of the problem \cite{Gilbert88}.
In this paper we shall consider two cases, depending on the type of flow shear necessary for the $B_r$ stretching.

In case (i) the helical flow is solid body for $r < 1$ and at rest for $r>1$ (the same conductivity is assumed in both domains). The flow shear is then infinite and localized at the discontinuity surface $r=1$. Gilbert \cite{Gilbert88} has shown that this dynamo is fast (positive growth rate in the limit of large magnetic Reynolds number) and thus very efficient to generate a helical magnetic field of same pitch as the flow.
In case (ii) the helical flow is continuous, and equal to zero for $r \ge 1$. 
The flow shear is then finite at any point.
Gilbert \cite{Gilbert88} has shown that such a smooth helical flow is a slow dynamo and that the dynamo action is localized at a resonant layer $r=r_0$ such that $0<r_0<1$. Contrary to case (i), having a conducting external medium is here not necessary.

In both cases some resonant conditions leading to dynamo action have been derived \cite{Roberts87,Gilbert88, Ruzmaikin88, Gilbert00, Gilbert03}. Such resonant conditions can be achieved by choosing an appropriate geometry of the helical flow, like changing its geometrical pitch.
They have been derived for a stationary flow $\textbf{U}(r,\theta,z)$ and can be generalized to a time-dependent flow of the form $\widetilde{\textbf{U}}(r,\theta,z)\cdot f(t)$ where $f(t)$ is a periodic function of time.
Now taking a flow composed of a mean part $\textbf{U}$ plus a fluctuating part
$\widetilde{\textbf{U}}\cdot f(t)$, we expect 
the dynamo threshold to depend on the geometry of each part of the flow accordingly
to the resonant condition of each of them and to the ratio of the intensities
$|\widetilde{\textbf{U}}| / |\textbf{U}|$.
However we shall see that in some cases even a small intensity of the fluctuating part may have a drastic influence.
The results also depend on the frequency of $f(t)$.

The Ponomarenko dynamo (case(i)) fluctuating periodically in time and with a fluctuation of infinitesimal magnitude  had already been the object of a perturbative approach \cite{Normand03}. Here we consider a fluctuation of arbitrary magnitude. Comparing our results for a small fluctuation magnitude with those obtained with the perturbative approach, we found significant differences. Then we realized that there was an error in the computation of the results published in \cite{Normand03}(though the perturbative development in itself is correct). In Appendix \ref{Corrigendum} we give a corrigendum of these results.

\section{Model}
We consider a dimensionless flow defined in cylindrical coordinates $(r, \theta, z)$ by 
\begin{equation}
\textbf{U} = (0, r \Omega(r,t), V(r,t)) \cdot  h(r) \quad
\mbox{with} \quad h(r) = \left\{ \begin{array}{l} 1, \quad r<1  \\
0, \quad r>1 \\
\end{array} \right.,
\label{vitesse}
\end{equation} 
corresponding to a helical flow in a cylindrical cavity which is infinite in the $z$-direction, the external medium being at rest. Each component, azimuthal and vertical, of the dimensionless velocity is defined as the sum of a stationary part and of a fluctuating part 
\begin{equation}
	\Omega (r,t) = \left(\overline{R}_m + \widetilde{R}_m f(t)\right)\xi(r), \quad
	V(r,t) = \left(\overline{R}_m \overline{\Gamma}+ \widetilde{R}_m \widetilde{\Gamma} f(t)\right)\zeta(r)
	\label{flow}
\end{equation}
where $\overline{R}_m$ and $\overline{\Gamma}$ (resp. $\widetilde{R}_m$ and $\widetilde{\Gamma}$) 
are the magnetic Reynolds number and a characteristic pitch of the stationary (resp. fluctuating) part
of the flow.
In what follows we consider a fluctuation periodic in time, in the form 
$f(t) = \cos (\omega_f t)$.
Depending on the radial profiles of the functions
 $\xi$ and $\zeta$ we determine two cases (i) solid body
 and (ii) smooth flow
\begin{eqnarray}
	\mbox{(i)} && \xi = \zeta = 1 \\
	\mbox{(ii)}&&\xi = 1 - r, \;\;\zeta = 1-r^2.
	\label{defradial}
\end{eqnarray}
We note here that the magnetic Reynolds numbers are defined with the maximum angular velocity
(either mean or fluctuating part) and the radius of the moving cylinder. 
Thinking of an experiment, it would not be sufficient to minimize the
magnitude of the azimuthal flow. In particular if $\overline{\Gamma}$ is large (considering a steady flow for simplicity), one would have to spend too many megawatts in forcing the $z$-velocity. 
Therefore the reader interested in linking our results to experiments should bear in mind that our magnetic Reynolds number is not totally adequate for it.
A better definition of the magnetic Reynolds number could be for example
$\hat{R}_m = \overline{R}_m \sqrt{1 + \overline{\Gamma}^2}$.
For a stationary flow of type (i), the minimum dynamo threshold $\hat{R}_m$ is obtained for
$\overline{\Gamma}=1.3$.\\

Both cases (i) and (ii) differ in the conductivity of the external medium $r > 1$.
In case (i) the magnetic generation being in a cylindrical layer in the neighbourhood of $r=1$,
a conducting external medium is necessary for dynamo action. For simplicity we choose the same conductivity as the inner fluid.
In the other hand, in case (ii) the magnetic generation being within the fluid, 
a conducting external medium is not necessary for dynamo action, thus we choose an isolating external medium. 
Though the choice of the conductivity of the external medium is far from being insignificant for a dynamo experiment \cite{Avalos03,Avalos05,Frick02,Ravelet05},
we expect that it does not change the overall meaning of the results given below.\\ 
We define the magnitude ratio of the fluctuation to the mean flow by
$\rho=\widetilde{R}_m/\overline{R}_m$. For $\rho=0$ there is no fluctuation and the dynamo threshold is given by $\overline{R}_m$. In the other hand for $\rho \gg 1$ the fluctuation dominates and the relevant quantity to 
determine the threshold is $\widetilde{R}_m = \rho \overline{R}_m$. The perturbative approach of Normand \cite{Normand03} correspond to $\rho\ll 1$.\\

The magnetic field must satisfy the induction equation
\begin{equation}
	\frac{\partial \textbf{B}}{\partial t} = \nabla \times (\textbf{U} \times \textbf{B}) + \nabla^2 \textbf{B}.
\label{induc}
\end{equation}
where the dimensionless time $t$ is given in units of the magnetic diffusion time, implying that the flow frequency $\omega_f$
is also a dimensionless quantity.
As the velocity does not depend on $\theta$ nor $z$,
each magnetic mode in $\theta$ and $z$ is independent from the others.
Therefore we can look for a solution in the form
\begin{equation}
\textbf{B}(r,t)= \exp i(m \theta + k z) \textbf{b}(r,t)
\label{defb}
\end{equation}
where $m$ and $k$ are the azimuthal and vertical wave numbers of the field.
The solenoidality of the field $\nabla \cdot \textbf{B} = 0$ then leads to 
\begin{equation}
	\frac{b_r}{r} + b_r' + i\frac{m}{r}b_{\theta} + ik b_z = 0.
	\label{divBzero}
\end{equation}

With the new variables $b^{\pm}= b_r \pm ib_{\theta}$,
the induction equation can be written in the form
\begin{equation}
\frac{\partial b^{\pm}}{\partial t} + [k^2 +
i (m \Omega + k V) h(r)]b^{\pm}= \pm \frac{i}{2}r \Omega' h(r)(b^+ + b^-) + {\cal L}^{\pm}b^{\pm} \;,
\label{inducadim}
\end{equation}
with
\begin{equation}
{\cal L}^{\pm} = \frac{\partial^2}{\partial r^2} + \frac{1}{r}\frac{\partial}{\partial r} - \frac{(m\pm 1)^2}{r^2},	
\end{equation}
except in case (ii) where in the external domain $r>1$, as it is non conducting, the induction equation takes the form
\begin{equation}
\left({\cal L}^{\pm} - k^2\right)b^{\pm} = 0.
\label{domextii}
\end{equation}
At the interface $r=1$, both $\textbf{B}$ 
and the $z$-component of the electric field $\textbf{E} = \nabla \times \textbf{B} - \textbf{U} \times \textbf{B} $ are continuous. The continuity of $B_r$ and $B_{\theta}$ imply that of $b^{\pm}$. The continuity of $\textbf{B}$ and (\ref{divBzero}) imply the continuity
of $b'_r$ which, combined with the continuity of $E_z$ implies
\begin{equation}
	[D b^{\pm}]_{1+}^{1-} \pm \frac{i\Omega_{r=1^-}}{2} (b^+ + b^-)_{r=1}=0
	\label{saut1}
\end{equation}
 with
$D= \partial / \partial r$ and $[h]_{1+}^{1-} = h_{(r=1-)} - h_{(r=1+)}$.
We note that in case (ii) as $\Omega_{r=1^-}=0$, (\ref{saut1}) implies the continuity of $D b^{\pm}$ at $r=1$.

In summary, we calculate for both cases (i) and (ii) the growth rate
\begin{equation}
	\gamma = \gamma(m,k,\overline{\Gamma},\widetilde{\Gamma},\overline{R}_m,\widetilde{R}_m,\omega_f)
\end{equation}
of the kinematic dynamo problem and look for the dynamo threshold (either $\overline{R}_m$ or $\widetilde{R}_m$) such that the real part $\Re\gamma$ of $\gamma$ 
is zero.
In our numerical simulations we shall take $m=1$ for it leads to the lowest
dynamo threshold.
\subsection{Case (i): Solid body flow}
In case (i) we set
\begin{eqnarray}
  m \Omega + k V = \overline{R}_m\overline{\mu} + \widetilde{R}_m\widetilde{\mu} f(t), \quad \mbox{with} \quad
	\overline{\mu} = m+k\overline{\Gamma} \quad \mbox{and} \quad \widetilde{\mu} = m+k\widetilde{\Gamma},
	\label{defmus}
\end{eqnarray}
and (\ref{inducadim}) changes into
\begin{equation}
\frac{\partial b^{\pm}}{\partial t} + [k^2 +
i (\overline{R}_m\overline{\mu}+\widetilde{R}_m \widetilde{\mu} f(t)) h(r)]b^{\pm}= {\cal L}^{\pm}b^{\pm} .
\label{inducadim2}
\end{equation}
For mathematical convenience, we take $\widetilde{\mu} = 0$ . 
Then the non stationary part of the velocity does not occur in 
(\ref{inducadim2}) any more. It occurs only in the expression of the boundary conditions 
(\ref{saut1}) that can be written in the form
\begin{equation}
	[D b^{\pm}]_{1+}^{1-} \pm \frac{i}{2}(\overline{R}_m + \widetilde{R}_mf(t)) (b^+ + b^-)_{r=1}=0.
	\label{saut}
\end{equation}
Taking $\widetilde{\mu} = 0$ corresponds to a pitch of the magnetic field  equal to the
pitch of the fluctuating part of the flow $-m/k=\widetilde{\Gamma}$.
In the other hand it is not necessarily equal to the pitch of the mean flow (except if $\overline{\Gamma}=\widetilde{\Gamma}$).
In addition we shall consider two situations depending on whether the mean flow is zero ($\overline{R}_m = 0$) or not. The method used to solve the equations (\ref{inducadim2}) and (\ref{saut}) is given in Appendix \ref{resolution i}.\\

At this stage we can make two remarks.
First, according to boundary layer theory results \cite{Roberts87,Gilbert88}
and for a stationary flow,
in the limit of large $\overline{R}_m$ the magnetic field
which has the highest growth rate satisfies $\overline{\mu} \approx 0$.
This resonant condition means that the pitch 
of the magnetic field is roughly equal to the pitch of the flow.
We shall see in section \ref{section:stationary} that this stays
true even at the dynamo threshold.
Though the case of a fluctuating flow of type $\widetilde{\textbf{U}}\cdot f(t)$
may be more complex with possibly skin effect, the resonant condition
is presumably analogous, writing $\widetilde{\mu} \approx 0$. This means that
setting $\widetilde{\mu} = 0$ implies that if the fluctuations are sufficiently large ($\rho \gg 1$), dynamo action is always possible. This is indeed what will be found in our results. In other words, setting $\widetilde{\mu} = 0$, we cannot 
tackle the situation of a stationary dynamo flow to which a fluctuation acting against the dynamo would be added. This aspect will be studied with the smooth flow (ii).

Our second remark is about the effect of a phase lag between the azimuthal and vertical components of the flow fluctuation. Though we did not study
the effect of an arbitrary phase lag we can predict the effect of 
an out-of-phase lag. This would correspond to take a negative value of $\widetilde{\Gamma}$.
Solving numerically the equations (\ref{inducadim2}) and (\ref{saut}) for the stationary flow and $m=1$, we find that dynamo action is possible only if $k \overline{\Gamma} <0 $. For the fluctuating flow with zero mean, $m=1$ and $\widetilde{\mu}=0$ necessarily implies that $k \widetilde{\Gamma} = -1 $.  
Let us now consider a flow containing both a stationary and a fluctuating part.
Setting $\widetilde{\Gamma} < 0$ necessarily implies that $k>0$. Then for $\overline{\Gamma} > 0$,  the stationary flow is not a dynamo.
Therefore in that case we expect the dynamo threshold to decrease for increasing $\rho$. 
For $\overline{\Gamma} <0$, together with $\widetilde{\Gamma} < 0$ and $k>0$, it is equivalent to take $\widetilde{\Gamma} > 0$ and $\overline{\Gamma} >0$ for $k<0$
and it is then covered by our subsequent results.

\subsection{Case (ii): Smooth flow}
For the case (ii) we can directly apply the resonant condition made up for a stationary flow  \cite{Gilbert88,Ruzmaikin88}, to the case of a fluctuating flow. For given $m$ and $k$, 
the magnetic field is generated in a resonant layer $r=r_0$ where the magnetic field lines are aligned with the shear and thus minimize the magnetic field diffusion.
This surface is determined by the following relation \cite{Gilbert88,Ruzmaikin88}
\begin{equation}
	m\Omega'(r_0) + k V'(r_0) = 0.
	\label{resonnance}
\end{equation}
The resonant condition is satisfied if the resonant surface is embedded within the fluid 
\begin{equation}
	0 < r_0 < 1.
	\label{resonantcond}
\end{equation}
As $\Omega$ and $V$ depend on time, this condition may only be satisfied at discrete times. This implies
successive periods of growth and damping, the dynamo threshold corresponding to a zero mean growth rate.
We can also define two distinct resonant surfaces $\overline{r}_0$ and $\widetilde{r}_0$ corresponding
to the mean and fluctuating part of the flow,
\begin{equation}
	m\overline{\Omega}'(\overline{r}_0) + k \overline{V}'(\overline{r}_0) = 0,
	\quad \quad \quad
	m\widetilde{\Omega}'(\widetilde{r}_0(t),t) + k \widetilde{V}'(\widetilde{r}_0(t),t) = 0
	\label{defr0}
\end{equation}
with appropriate definition of $\overline{\Omega}, \overline{V}, \widetilde{\Omega}$
and $\widetilde{V}$. In addition, if $\widetilde{\Omega}$ and $\widetilde{V}$
have the same time dependency, as in (\ref{flow}), then $\widetilde{r}_0$ becomes time independent.
Then we can predict two different behaviours of the dynamo threshold versus the fluctuation rate
$\rho=\widetilde{R}_m/\overline{R}_m$. If $0 < \overline{r}_0 < 1$ and $\widetilde{r}_0 >1$
then the dynamo threshold will increase with $\rho$. In this case the fluctuation is harmful
to dynamo action. In the other hand if $0 < \widetilde{r}_0 < 1$ then the dynamo threshold will decrease
with $\rho$.

From the definitions (\ref{defr0}) and for a flow defined by
(\ref{vitesse}), (\ref{flow}) and (\ref{defradial})
we have 
\begin{equation}
	\overline{r}_0 = -(m/k)/(2\overline{\Gamma}) \quad \quad \mbox{and}
	\quad \quad 
	\widetilde{r}_0 = -(m/k)/(2\widetilde{\Gamma}).
	\label{defdesr0}
\end{equation}
For $m=1$ and $k<0$, taking $\widetilde{\Gamma}<0$
implies $\widetilde{r}_0 < 0$ and then the impossibility
of dynamo action for the fluctuating part of the flow.
Therefore, we expect that the addition of a fluctuating flow with an out-of-phase lag between its vertical and azimuthal components will necessarily be harmful to dynamo action. This will be confirmed numerically in section \ref{sec:moynonul}. 

To solve (\ref{inducadim}), (\ref{domextii}) and (\ref{saut1}),
we used a Galerkin approximation method 
in which the trial and weighting functions 
are chosen in such a way that the resolution of the induction equation is reduced to the conducting domain
$r \le 1$ \cite{Marie06}.
The method of resolution is given in Appendix \ref{resolution ii}.
For the time resolution we used a Runge-Kutta scheme of order 4.

\section{Results}
\subsection{Stationary flow ($\widetilde{R}_m=0$)}
\label{section:stationary}
\begin{figure}
\begin{tabular}{@{\hspace{-2cm}}l}
    \includegraphics[width=1.2\textwidth]{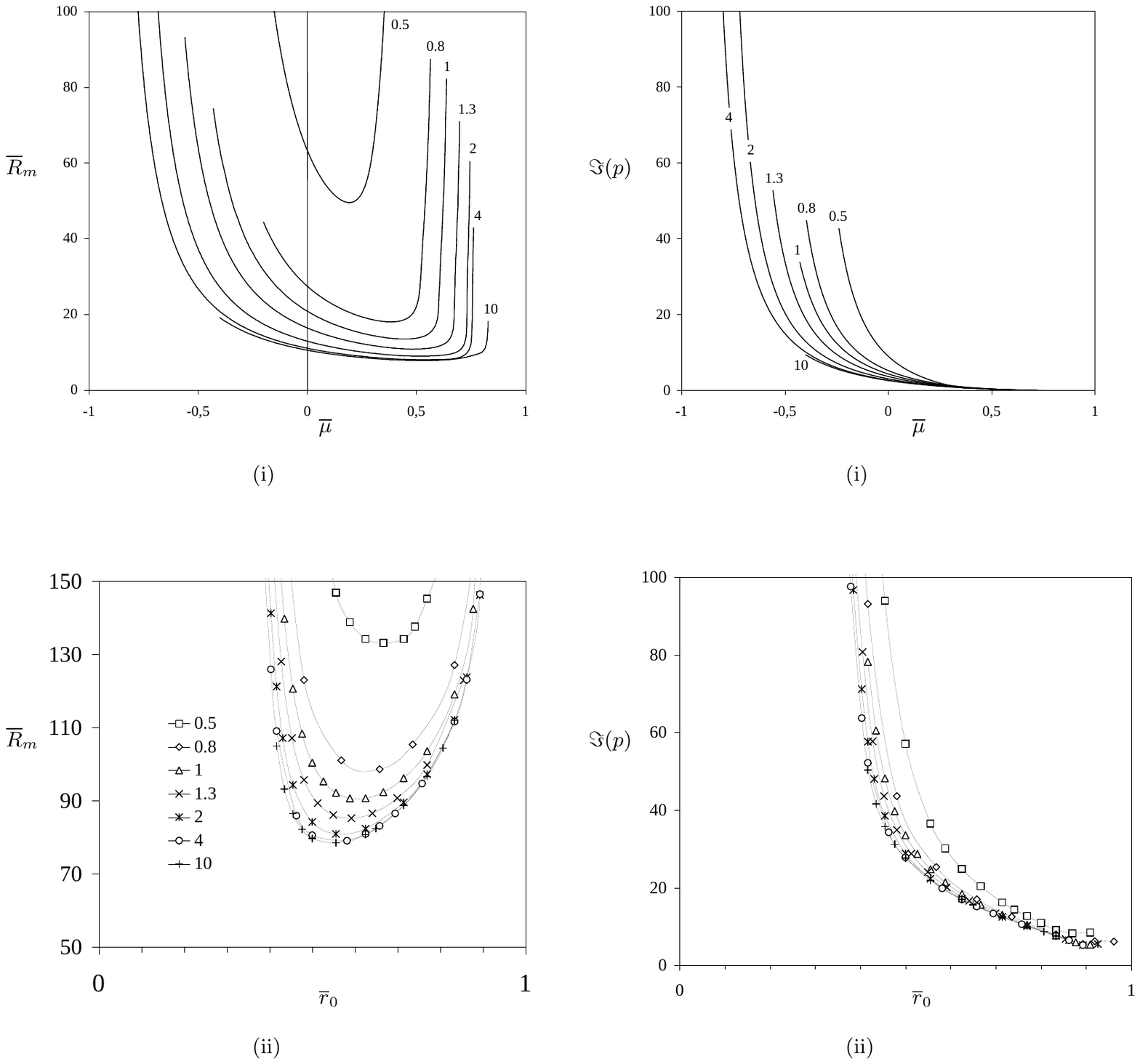}\\*[-13cm]
    \end{tabular}
\caption{The dynamo threshold $\overline{R}_m$ (left column) and $\Im (\gamma)$ (right column) versus (i) $\overline{\mu}$, (ii) $\overline{r}_0$, for the stationary case,
$m=1$ and $\overline{\Gamma} = 0.5; 0.8; 1; 1.3; 2; 4; 10$ .}
\label{figstat}
\end{figure}

We solve
\begin{equation}
	\Re \gamma(m=1,\;\;k,\;\;\overline{\Gamma},\;\;0,\;\;\overline{R}_m,\;\;0,\;\;0)=0
\end{equation}
with $ k=(\overline{\mu} -1)/\overline{\Gamma}$ for case (i) and $k=-1/ (2 \overline{r}_0 \overline{\Gamma})$
for case (ii).
In figure \ref{figstat} the threshold $\overline{R}_m$ and the field frequency $\Im (\gamma)$ are plotted
versus $\overline{\mu}$ (resp. $\overline{r}_0$) for case (i) (resp. (ii)), and
for different values of $\overline{\Gamma}$.
Though we do not know how these curves asymptote, and though the
range of $\overline{\mu}$ (resp. $\overline{r}_0$) for which dynamo action occurs changes with $\overline{\Gamma}$,
it is likely that the resonant condition  $|\overline{\mu}|<1$ (resp. $0<\overline{r}_0<1$) is fulfilled for the
range of $\overline{\Gamma}$ corresponding to a dynamo experiment ($\overline{\Gamma} \approx 1$).
In case (i) the dispersion relation (\ref{systeme2}) 
in Appendix \ref{resolution i}
becomes ${\cal F}_0 = 0$. 

\subsection{Periodic flow with zero mean ($\overline{R}_m = 0$)}
\begin{figure}
  \begin{tabular}{@{\hspace{2cm}}c@{\hspace{-1cm}}c@{\hspace{0cm}}l@{}}
    \raisebox{7.5cm}{$\widetilde{R}_m$} 
    &
    \includegraphics[width=1\textwidth]{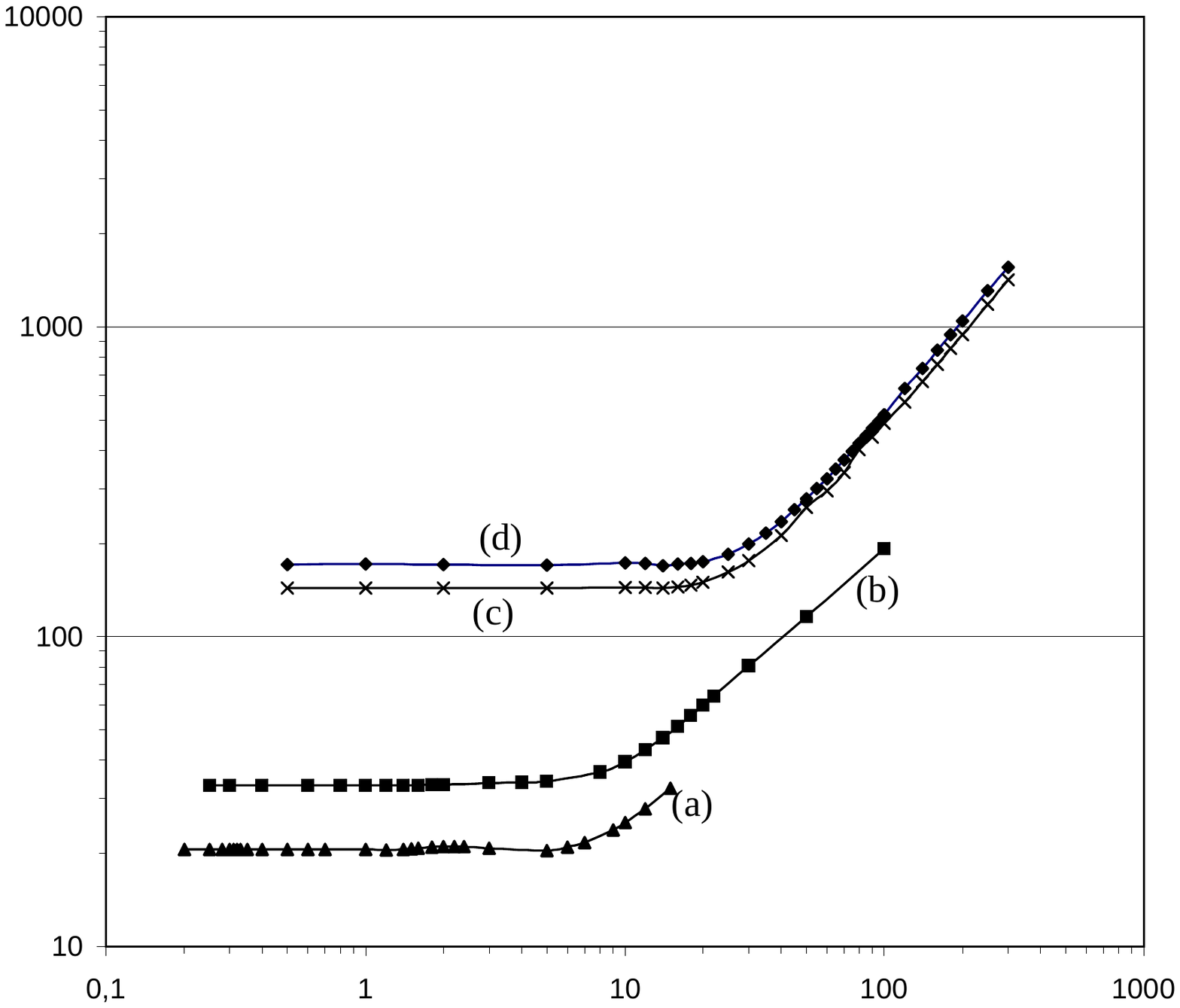}
    \\*[-1.8cm]
    &\hspace{3cm}$\omega_f$\\*[0cm]
  \end{tabular}
\caption{Dynamo threshold $\widetilde{R}_m$ versus $\omega_f$ for case (i)  with $\widetilde{\mu}=0$ and (a) $\widetilde{\Gamma}=1.78$, (b) $\widetilde{\Gamma}=1$;
for case (ii)  with $\widetilde{r}_0=0.5$ and (c) $\widetilde{\Gamma}=2$, (d) $\widetilde{\Gamma}=1$.}
\label{figoscillant}
\end{figure}

\begin{figure}
\begin{tabular}{@{\hspace{-2cm}}l}
    \includegraphics[width=1.2\textwidth]{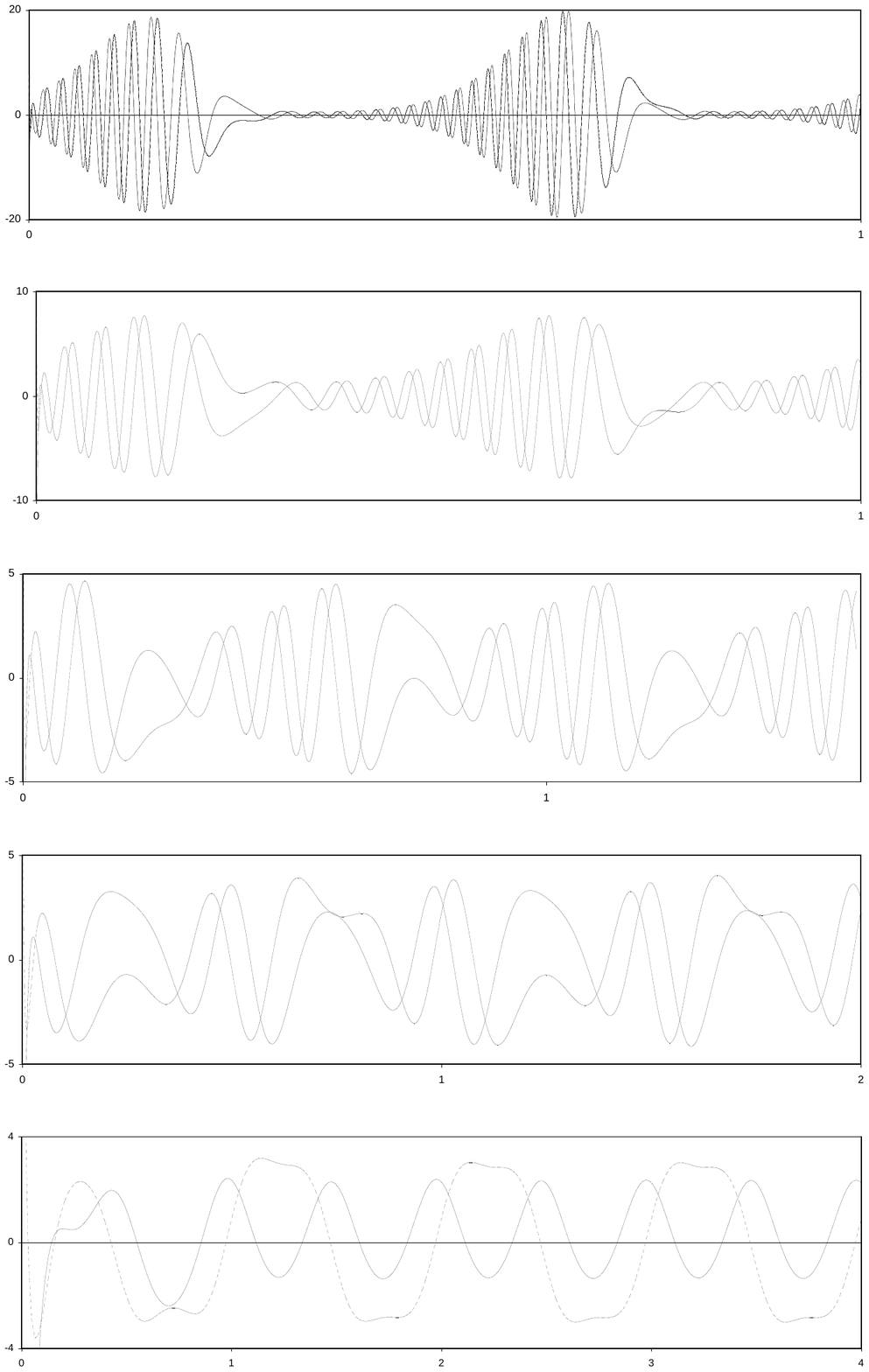}\\*[-7cm]
    \end{tabular}
    \caption{Time evolution of $\Re(b^-)$ (solid lines) and $\Im(b^-)$ (dotted lines) for several values of $\omega_f$ (from top to bottom $\omega_f=1; 2; 5; 10; 100$), for case (ii) with $\widetilde{\Gamma}= 1$. Time unity corresponds here to $2\pi/\omega_f$. }
\label{bmoinsfdet}
\end{figure}

We solve
\begin{equation}
	\Re \gamma(m=1,\;\;k,\;\;0,\;\;\widetilde{\Gamma},\;\;0,\;\;\widetilde{R}_m,\;\;\omega_f) =0.
\end{equation}
In figure \ref{figoscillant} the threshold $\widetilde{R}_m$ is plotted versus $\omega_f$
for both cases (i) and (ii).
In both cases we take $\widetilde{\mu}=0$ corresponding to $k=-1/\widetilde{\Gamma}$. For the case (ii) it
implies from (\ref{defdesr0}) that
$\widetilde{r}_0=1/2$, meaning that the resonant surface is embedded in the fluid and then
favourable to dynamo action.
In each case (i) $\widetilde{\Gamma}= 1; 1.78$ and (ii) $\widetilde{\Gamma}= 1; 2$, we observe two regimes,
one at low frequencies for which the threshold does not depend on $\omega_f$ 
and the other at high frequencies for which the threshold behaves like
$\widetilde{R}_m \propto \omega_f^{3/4}$.\\
To understand the existence of these two regimes, we pay attention to the time evolution of the magnetic field for different frequencies $\omega_f$.
In figure \ref{bmoinsfdet}, the time evolution of $b^-$ (real and imaginary parts) for case (ii) $\widetilde{\Gamma}= 1$ (case (c) in figure \ref{figoscillant}) is plotted for several frequencies $\omega_f$.
\subsubsection{Low frequency regime}
For low frequencies ($\omega_f=1$), 
we observe two time scales : periodic phases of growth and decrease of the field,  with a time scale equal to the period of the flow as expected by Floquet's theory. In addition the field has an eigen-frequency much higher than $\omega_f$.
In fact the slow phases of growth and decrease seem to occur every half period of the flow.
This can be understood from the following remarks. \\
First of all the growth (or decrease) of the field does not depend on the sign of the flow.
Indeed, from (\ref{inducadim}), we show that if $b^{\pm}(m,k)$ is a solution for ($\Omega$, $V$), then its complex conjugate
${b^{\pm}}^*(m,k)$ is a solution for ($-\Omega$, $-V$). Therefore we have $b^{\pm}(t+T/2) = {b^{\pm}}^*(t)$
where $T=2\pi/\omega_f$ is the period of the flow.
Now from Floquet's theory (see Appendix \ref{resolution i}),
we may write $\textbf{b}(r,t)$ in the form $\textbf{b}(r,\tau) \exp(\gamma t)$,
with $\textbf{b}(r,\tau)$ $2 \pi$-periodic in $\tau=\omega_f t$. This implies
that changing ($\Omega$, $V$) in ($-\Omega$, $-V$) changes the sign of $\Im(\gamma)$.
This is consistent with the fact that for given $m$ and $k$, 
the direction of propagation of $\textbf{B}$ changes with the direction of the flow.
Therefore changing the sign of the flow changes the sign of propagation of the field but
does not change the magnetic energy, neither the dynamo threshold $\widetilde{R}_m$ which are then identical from one half-period of the flow to another.
This means that the dynamo threshold does not change if we consider $f(t) = |\cos(\omega_f t)|$ instead of $\cos(\omega_f t)$. It is then sufficient to concentrate on one half-period of the flow like for example $[\frac{\pi}{2 \omega_f}, \frac{3\pi}{2 \omega_f}]$ (modulo $\pi$). \\
The second remark uses the fact that the flow geometry that we consider does not change in time (only the flow magnitude changes).
For such a geometry we can calculate the dynamo threshold $\overline{R}_m$ corresponding to the stationary case.
Then coming back to the fluctuating flow, we understand that $\widetilde{R}_m|f(t)| > \overline{R}_m$
(resp. $\widetilde{R}_m|f(t)| < \overline{R}_m$) corresponds to a growing (resp. decreasing) phase of the field. Assuming that the dynamo threshold $\widetilde{R}_m$ is given by the time average
$<.>$ of the flow magnitude leads to the following estimation for $\widetilde{R}_m$
\begin{equation}
	\widetilde{R}_m \approx \frac{\pi}{2}\overline{R}_m
\end{equation}
as  $<|\cos(\omega_f t)|> = 2 / \pi$.
For the three cases (a), (b) and (c) in figure \ref{figoscillant} 
we give in table \ref{table:comparaison} the ratio $2\widetilde{R}_m / \pi\overline{R}_m$ which is found to be always close to unity. 
\begin{table} 
\begin{center}
\begin{tabular}{|c|c|c|c|c|l|} \hline
   &$\widetilde{R}_m$& $\overline{R}_m$ & $\frac{2\widetilde{R}_m}{\pi\overline{R}_m}$& $\overline{\omega}$ \\ \hline
   (a) &21&13&1.03&4.4 \\ \hline
   (b) &33&21&1 &3.1\\ \hline
   (c) &143&84&1.08&28.8\\ \hline
   (d) &170&100&1.08&33\\ \hline
\end{tabular}
  \label{table:comparaison}
  \end{center}
    \caption{see in the text.}
\end{table}
In this interpretation of the results the frequency $\omega_f$ does not appear, provided
that it is sufficiently weak in order that the successive phases of growth and decrease have sufficient time to occur. This can explain why for low frequencies in figure \ref{figoscillant} the dynamo threshold $\widetilde{R}_m$ does not depend on $\omega_f$.

Finally the frequencies $\overline{\omega}$ of the stationary case for $\overline{\Gamma} = \widetilde{\Gamma}$
are also reported in table \ref{table:comparaison}.
For a geometry identical to case (c), we find, in the stationary case, $\overline{\omega} = 33$
which indeed corresponds to the eigen-frequency of the field occurring in figure \ref{bmoinsfdet}
for $\omega_f=1$.
The previous remarks assume that the flow frequency is sufficiently small compared to the eigen-frequency of the field, in order to have successive phases of growth and decrease of the field.
We can check that the values of $\overline{\omega}$ given in table \ref{table:comparaison} are indeed 
reasonable estimations of the transition frequencies between the low and high frequency regimes in figure \ref{figoscillant}.

\subsubsection{High frequency regime}
In case (ii) and for high frequencies (figure \ref{bmoinsfdet}, $\omega_f=100$), the signal 
is made of harmonics without growing nor decreasing phases.
We note that the eigen-frequency of the real and imaginary parts of $b^-$ are different,
the one being twice the other.

In case (i), relying on the resolution of equations (\ref{inducadim2}) and (\ref{saut}) given in Appendix
\ref{resolution i}, we can show that 
$\widetilde{R}_m \propto \omega_f^{3/4}$. We also find that some double frequency as found in figure \ref{bmoinsfdet}
for the case (ii) can emerge from an approximate $3 \times 3$ matrix system. As these developments necessitate the notations introduced in Appendix \ref{resolution i}, they are postponed in Appendix \ref{highfrequency}.

\subsubsection{Further comments about the ability for fluctuating flows to sustain dynamo action}
We found and explained how a fluctuating flow (zero mean) can act as a dynamo.
We also understood why the dynamo threshold for a fluctuating flow 
is higher than that for a stationary flow with the same geometry. It is
because the time-average of the velocity norm of the fluctuating flow is on the mean lower than that of the stationary flow.
This can be compensated with other definitions of the magnetic Reynolds number.
Our definition is based on $\max_t |\Omega(r,t)|$. An other definition based on $<|\Omega(r,t)|>_t$
would exactly compensate the difference.

Recently a controversy appeared about the difficulty for a fluctuating flow (zero mean) to sustain dynamo action at low $P_m$ \cite{Schekochihin04},
whereas a mean flow (non-zero time average) exhibits a finite threshold
at low $P_m$ \cite{Ponty07,Laval06} (the magnetic Prandtl number, $P_m$,
being defined as the ratio of the viscosity to the diffusivity of the fluid).
This issue is important not only for dynamo experiments but also for
natural objects like the Earth's inner-core or the solar convective zone in which the 
electro-conducting fluid is characterized by a low $P_m$.
Though we did not study this problem, our results suggest that the dynamo threshold should not
be much different between fluctuating and mean flows,
provided an appropriate definition of the magnetic Reynolds number is taken.
In that case why does it seem so difficult to sustain dynamo action
at low $P_m$ for a fluctuating flow \cite{Schekochihin04} whereas it
seems much easier for a mean flow \cite{Ponty07,Laval06} ?

In the simulations with a mean flow \cite{Ponty07,Laval06}, two dynamo regimes have been found,
the one with a threshold much lower than the other. In the lowest threshold regime, 
the magnetic field is generated at some infinite scale in two directions \cite{Ponty07}. 
There is then an infinite scale separation
between the magnetic and the velocity field, and the dynamo action 
is probably of mean-field type
and might be understood in terms of $\alpha$-effect, $\beta$-effect, etc.
In that case, removing the periodic boundary conditions would cancel the scale separation and
imply the loss of the dynamo action.
In the highest threshold regime,
the magnetic field is generated at a scale similar to the flow scale, the periodic boundary conditions are forgotten
and the dynamo action can not be 
understood in
terms of an $\alpha$-effect anymore. 
In order to compare the mean flow results \cite{Ponty07} with those for a fluctuating flow
\cite{Schekochihin04}
we have to consider only the highest threshold regime in \cite{Ponty07},
the lowest one relying on mean-field dynamo processes
due to the periodic boundary conditions
and which are absent in the fluctuating flow calculations \cite{Schekochihin04}. 

Now when comparing the threshold of
the highest threshold regime for a mean flow  with the threshold obtained for a fluctuating flow
and with appropriate definitions of $R_m$,
a strong difference remains at low $P_m$ \cite{Iskakov07}.
A speculation made by Schekochihin \textit{et al.} \cite{Schekochihin07} is that the
highest threshold regime obtained for the mean flow at low $P_m$ \cite{Ponty07} would correspond in fact to the large $P_m$ results for the fluctuating flow in \cite{Iskakov07}.
Their arguments rely on the fact 
that the mean flow in \cite{Ponty07} is peaked at large scale and so 
spatially smooth for the generated magnetic field.
It would then belongs to the same class as the large-$P_m$ fluctuation dynamo.
Both dynamo thresholds are found to be similar indeed, and thus the discrepancy vanishes.

Though the helical flow that we consider here is noticeably different
(no chaotic trajectories) it may have some consistency with the simulations at large $P_m$
mentioned above and at least supports the speculation by Schekochihin \textit{et al.} \cite{Schekochihin07}.

\subsection{Periodic flow with non zero mean}
\label{sec:moynonul}
We are now interested in the case where both $\widetilde{R}_m \ne 0$ and $\overline{R}_m \ne 0$.
The flow is then the sum of a non zero mean part and a fluctuating part. We have considered two approaches depending on which part of the flow geometry is fixed, either the mean or the fluctuating part.
\subsubsection{$\overline{\Gamma}=1$}
\begin{figure} 
\begin{tabular}{@{\hspace{-2cm}}l}
    \includegraphics[width=1.15\textwidth]{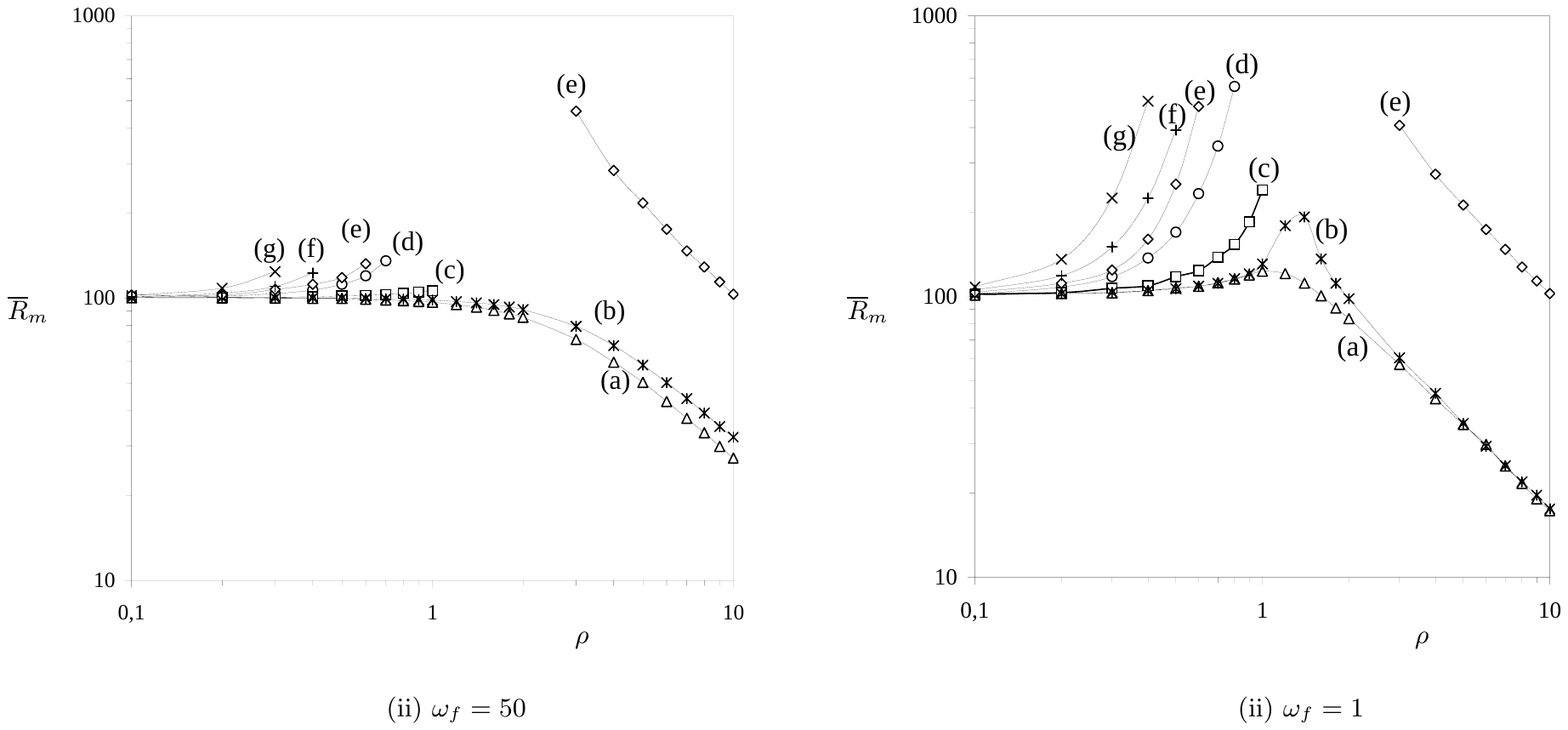}\\*[-19cm]
    \end{tabular}
\caption{Dynamo threshold $\overline{R}_m$ versus $\rho$ for case (ii), for two frequencies $\omega_f=50$ and $\omega_f=1$ and 
$\overline{r}_0=1/2$ ($\overline{\Gamma}=1$, $m=1$, $k=-1$). 
The different curves correspond to $\widetilde{r}_0=$ (a) 1/2; (b) 2/3; (c) 1; (d) $\infty$; (e) 1/4;
(f) -1; (g) -1/2 ($\widetilde{\Gamma} =$ (a) 1; (b) 0.75; (c) 0.5; (d) 0; (e) 2; (f) -0.5; (g) -1).}
\label{Rmper}
\end{figure}

Here we fix $\overline{\Gamma}=1$, $m=1$ and $k=-1$ and vary $\widetilde{\Gamma}$,
$\rho$ and $\omega_f$ for the case (ii).
Then we solve the equation
\begin{equation}
	\Re \gamma(m=1,\;\;k=-1,\;\;\overline{\Gamma}=1,\;\;\widetilde{\Gamma}=1/2\widetilde{r}_0,\;\;\overline{R}_m,\;\;\widetilde{R}_m=\rho \overline{R}_m ,\;\;\omega_f) =0
\end{equation}
to plot $\overline{R}_m$  as a function of $\rho$ in figure \ref{Rmper}
for values of $\widetilde{r}_0$ and $\omega_f$.
From (\ref{defdesr0}) we have $\overline{r}_0=1/2$ which corresponds to a mean flow geometry with a dynamo threshold about 100.
The curves are plotted for several values of  $\widetilde{\Gamma}$ 
leading to values of  $\widetilde{r}_0$ not necessarily between $0$ and $1$. 
We consider two fluctuation frequencies $\omega_f=1$ and $\omega_f=50$. 
We find that the dynamo threshold $\overline{R}_m$ 
increases asymptotically with $\rho$ unless the resonant condition $0<\widetilde{r}_0<1$ is satisfied, here the curves (a), (b) and (e). 
For these three curves we checked that in the limit of large $\rho$,
$\overline{R}_m = O(\rho^{-1})$. For $\widetilde{r}_0 = 1/4$ (curve (e)) and for $\rho \approx 1$ we do not know if a dynamo threshold exists.\\
\subsubsection{$\widetilde{\Gamma}=1$}
\begin{figure} 
\begin{tabular}{@{\hspace{-2cm}}l}
    \includegraphics[width=1.15\textwidth]{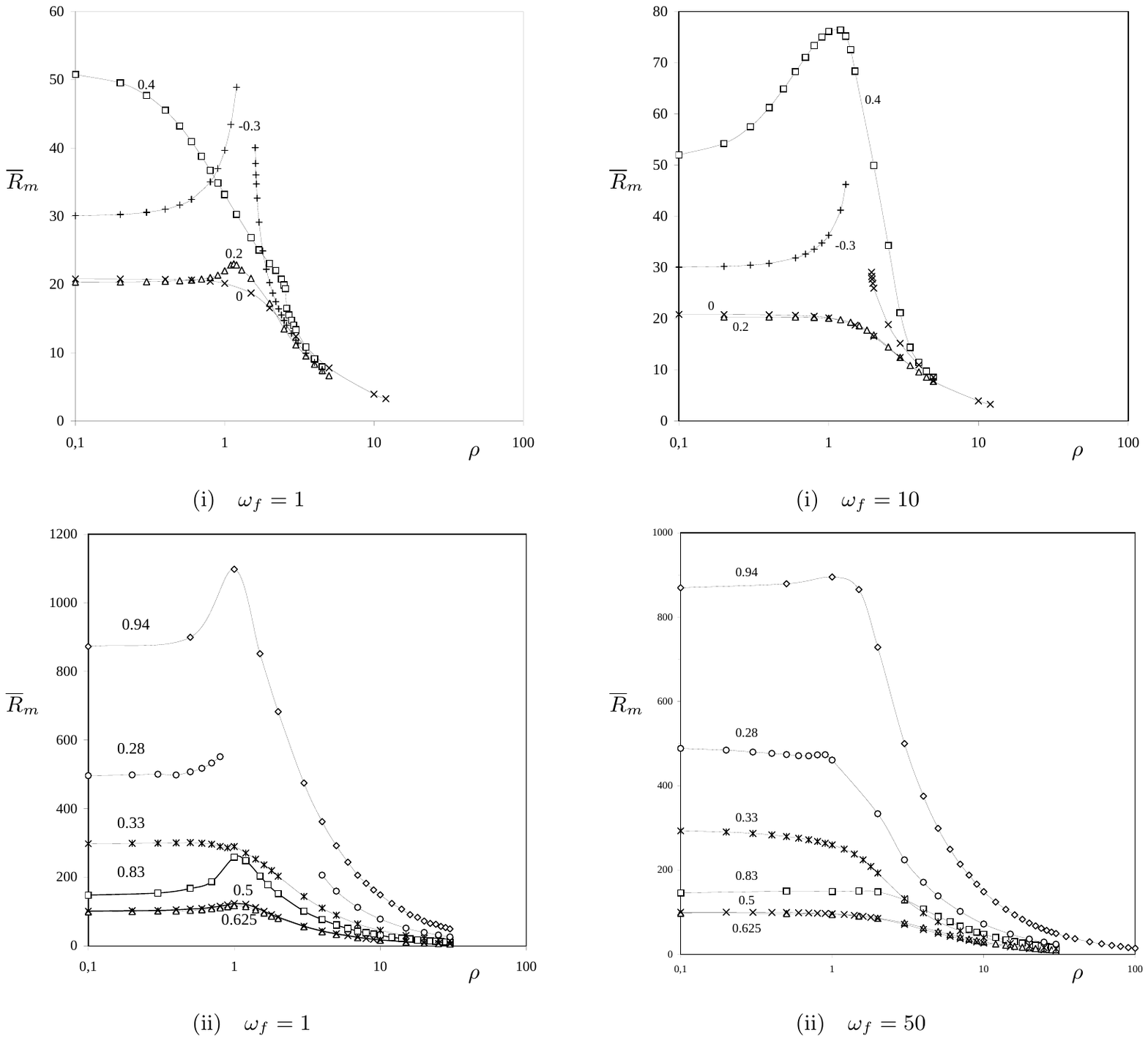}\\*[-13cm]
    \end{tabular}
\caption{The dynamo threshold $\overline{R}_m$ versus $\rho$ for $k=-1$, $m=1$ and $\widetilde{\Gamma}=1$ ($\widetilde{\mu}=0$ in case (i) and $\widetilde{r}_0=0.5$ in case (ii)) and $\omega_f=1, 10$ or 50. The labels correspond to $\overline{\mu}$ in case (i)
and $\overline{r}_0$ in case (ii).}
\label{figperiod}
\end{figure}

Here we fix  $\widetilde{\Gamma}=1$, $m=1$ and $k=-1$ and vary $\overline{\Gamma}$,
$\rho$ and $\omega_f$.
We then solve the equation
\begin{equation}
	\Re \gamma(m=1,\;\;k=-1,\;\;\overline{\Gamma},\;\;\widetilde{\Gamma},\;\;\overline{R}_m,\;\;\widetilde{R}_m=\rho \overline{R}_m ,\;\;\omega_f) =0
\end{equation}
with $\overline{\Gamma}=1-\overline{\mu}$ in case (i) and $\overline{\Gamma}=1/2\overline{r}_0$ in case (ii).
In figure \ref{figperiod},
$\overline{R}_m$  is plotted versus $\rho$ 
for values of $\overline{\mu}$ (resp. $\overline{r}_0$) in case (i) (resp. (ii)) and $\omega_f$.
Taking $\widetilde{\Gamma}=1$, $m=1$ and $k=-1$ implies $\widetilde{\mu}=0$ in case (i)
and $\widetilde{r}_0=0.5$ in case (ii). In both cases (i) and (ii) the fluctuating part of the flow satisfies the resonant condition for which dynamo action is possible. This
implies that $\overline{R}_m$ should scale as $O(\rho^{-1})$ provided that $\rho$ is sufficiently large. In each case we consider two flow frequencies,
$\omega_f=1$ and $\omega_f=10$ for case (i), $\omega_f=1$ and $\omega_f=50$ for case (ii). The curves are plotted for different values of $\overline{\Gamma}$ corresponding to
$|\overline{\mu}|<1$ for case (i) and  $0<\overline{r}_0<1$ for case (ii). 
For large $\rho$ we checked that $\overline{R}_m = O(\rho^{-1})$.
The main differences between the curves is that $\overline{R}_m$ versus $\rho$ may decrease monotonically or not.
In particular in case (i) for $\overline{\mu}=0.4$, $\overline{R}_m$ decreases by 40\% when $\rho$ goes from 0 to 1 showing that even a small fluctuation can strongly decrease the dynamo threshold. In most of the curves there is a bump for $\rho$ around unity showing a strong increase of the threshold before the final decrease at larger $\rho$.
\subsubsection{$\overline{\Gamma}=\widetilde{\Gamma}=1$}
\begin{figure}
  \begin{tabular}{@{\hspace{2cm}}c@{\hspace{-1cm}}c@{\hspace{0cm}}l@{}}
    \raisebox{7.5cm}{$\overline{R}_m$} 
    &
    \includegraphics[width=1\textwidth]{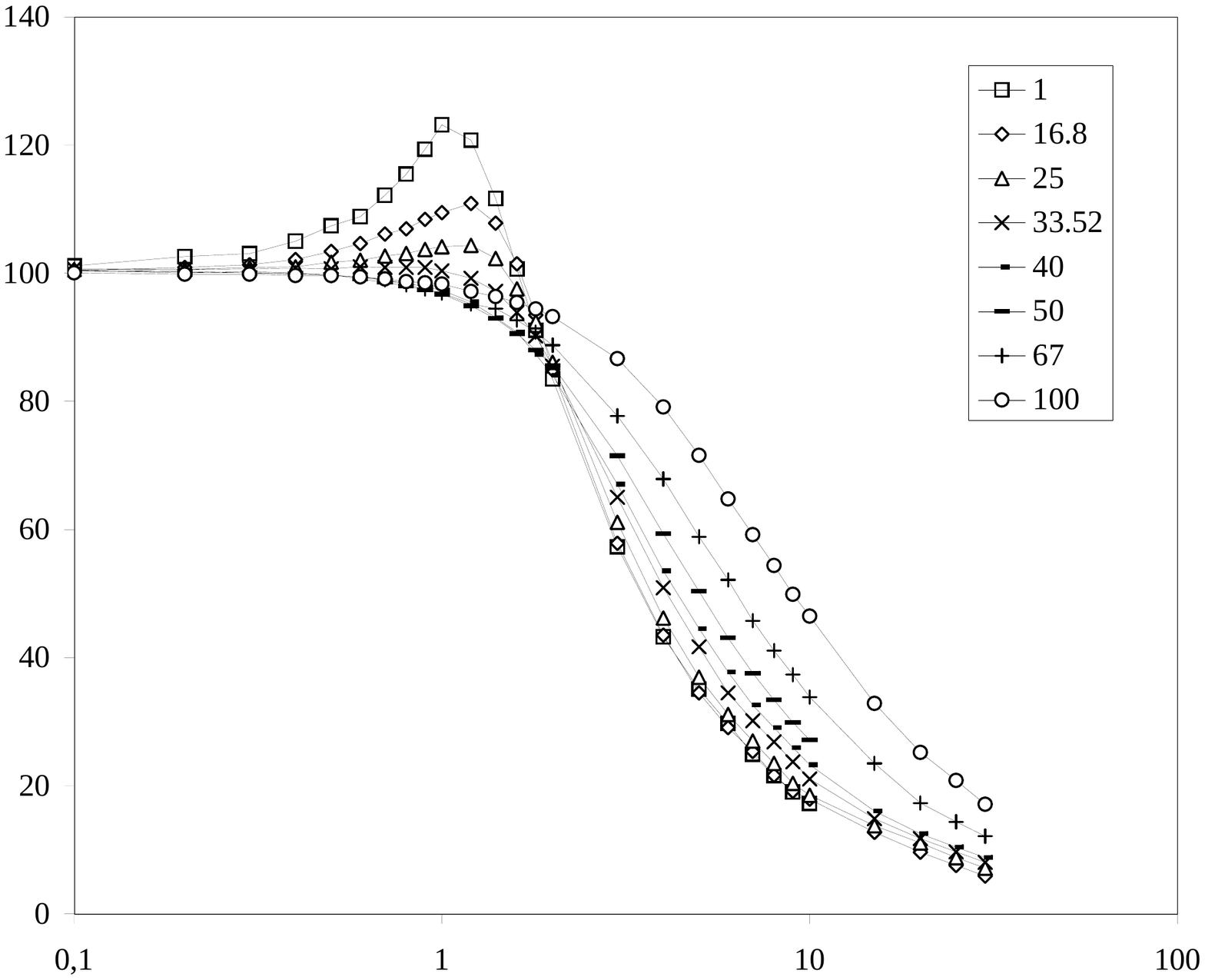}
    \\*[-1.8cm]
    &\hspace{3cm}$\rho$\\*[0cm]
  \end{tabular}
\caption{Dynamo threshold $\overline{R}_m$ versus $\rho$ for $\overline{r}_0=\widetilde{r}_0=0.5$ ($\overline{\Gamma} = \widetilde{\Gamma} = 1$). The labels correspond to different values of
$\omega_f$. The eigen-frequency for $\rho=0$ is $\overline{\omega}=33$.}
\label{fig:resonance}
\end{figure}

Here we fix  $\overline{\Gamma}=\widetilde{\Gamma}=1$, $m=1$ and $k=-1$
for the case (ii) and vary $\omega_f$ and $\rho$.
We then solve the equation
\begin{equation}
	\Re \gamma(m=1,\;\;k=-1,\;\;\overline{\Gamma}=1,\;\;\widetilde{\Gamma}=1,\;\;\overline{R}_m,\;\;\widetilde{R}_m=\rho \overline{R}_m ,\;\;\omega_f) =0
\end{equation}
to plot $\overline{R}_m$ versus $\rho$ in figure \ref{fig:resonance} for various frequencies $\omega_f$. 
Taking $\overline{\Gamma}=\widetilde{\Gamma}=1$, $m=1$ and $k=-1$
implies $\overline{r}_0=\widetilde{r}_0=0.5$
For $\rho$ larger than 1,  $\overline{R}_m$ decreases as $O(\rho^{-1})$ as mentioned earlier. For $\rho$ smaller than unity, $\overline{R}_m$ decreases versus $\rho$ monotonically only if $\omega_f$ is large enough. In fact the transition value of $\omega_f$ above which $\overline{R}_m$ decreases monotonically is exactly the field frequency $\overline{\omega}$ (here $\overline{\omega} = 33$) corresponding to $\rho=0$  . This shows that a fluctuation of small intensity ($\rho \le 1$) helps the dynamo action only if its frequency is sufficiently high. This is shown in Appendix \ref{highfrequency2} for the case (i).
Though, the frequency above which a small fluctuation intensity helps the dynamo may be much larger than $\overline{\omega}$.
For example in case (i) 
for $\overline{\mu}=0.4$ and $\widetilde{\mu}=0$ represented
in figure \ref{figperiod}, we have $\overline{\omega} = 0.51$. For $\omega_f=1$
small fluctuation helps, for $\omega_f=10$ they do not help, and for 
higher frequencies they help again.
\section{Discussion}
In this paper we studied the modification of the dynamo threshold of a stationary helical flow by the addition of a large scale helical fluctuation.
We extended a previous asymptotic study \cite{Normand03} to the case of
a fluctuation of arbitrary intensity (controlled by the parameter $\rho$).
We knew from previous studies \cite{Gilbert88,Ruzmaikin88} that the dynamo efficiency of a helical flow is characterized by some resonant condition at large $R_m$. 
First we verified numerically that such resonant condition holds at lower $R_m$
corresponding to the dynamo threshold,
for both a stationary and a fluctuating (no mean) helical flow.
Then for a helical flow made of a mean part plus a fluctuating part
we showed that, in the asymptotic cases $\rho \ll 1$ (dominating mean) and $\rho \gg 1$ (dominating fluctuation), it is naturally the resonant condition of the mean (for the first case) or the fluctuating (for the second case) part of the flow which governs the dynamo efficiency and then the dynamo threshold. 
In between, for $\rho$ of order unity and if the resonant condition of each flow part (mean and fluctuating) is satisfied, the threshold first increases with $\rho$ before reaching an asymptotic behaviour in $O(\rho^{-1})$.
However there is no systematic behaviour as depicted in figure \ref{figperiod} (i) for $\omega_f=1$ and $\overline{\mu}=0.4$, in which a threshold decrease of $40 \%$ is obtained between $\rho=0$ and $\rho=1$.
If the fluctuation part of the flow does not satisfy the resonant condition, then the dynamo threshold increases drastically with $\rho$.

Contrary to the case of a cellular flow \cite{Petrelis06}, there is no systematic effect
of the phase lag between the different components of the helical flow. For the helical flow geometry it may imply an increase or a decrease of the dynamo threshold, depending how it changes the resonant condition mentioned above.

There is some similarity between our results and those obtained for a noisy (instead of periodic) fluctuation \cite{Leprovost05}.
In particular in \cite{Leprovost05} it was found that increasing the noise level the threshold first increases due to geometrical effects of the magnetic field lines and then decreases at larger noise. This could explain why at $\rho \approx 1$ we generally obtain a maximum of the dynamo threshold. 

Finally these results show that the optimization of a dynamo experiment depends not only on the mean part of the flow but also on its non stationary large scale part. If the fluctuation is not optimized then the threshold may increase drastically with (even small) $\rho$, ruling out any hope of producing dynamo action. In addition, even if the fluctuation is optimized (resonant condition satisfied by the fluctuation),
our results suggest that there is generally some increase of the dynamo threshold with $\rho$ 
when $\rho \le 1$. If the geometry of the fluctuation is identical to that of the mean part of the flow, there can be some slight decrease of the threshold at high frequencies but this decrease is rather small. When $\rho > 1$ the dynamo threshold decreases as $O(\rho^{-1})$ which at first sight seems interesting.
However we have to keep in mind that as soon as $\rho >1$ the driving power spent to maintain the fluctuation is larger than that to maintain the mean flow. Then the relevant dynamo threshold is not $\overline{R}_m$ any more, but $\widetilde{R}_m =  \rho \overline{R}_m$ instead. In addition monitoring large scale fluctuations in an experiment may not always be possible, especially if they occur from flow destabilisation.
 In that case it is better to try cancelling them
 as was done in the VKS experiment in which
an azimuthal belt has been added \cite{Monchaux06}.

\section{Acknowledgments} 
We acknowledge B. Dubrulle, F. P\'etr\'elis, R. Stepanov and A. Gilbert for fruitful discussions.
\section{Appendix}
\subsection{Resolution of equations (\ref{inducadim2}) and (\ref{saut}) for the case (i): solid body flow}
\label{resolution i}
As $f(t)$ is time-periodic of period $2 \pi / \omega_f$, 
we look for $\textbf{b}(r,t)$ in the form $\textbf{b}(r,\tau) \exp(\gamma t)$
with $\textbf{b}(r,\tau)$ being $2 \pi$-periodic in $\tau=\omega_f t$. 
Thus we look for the functions $b^{\pm}(r,\tau)$
in the form
\begin{equation}
b^{\pm}(r,\tau) = \sum b^{\pm}_n(r) \exp (in\tau)
\end{equation}
where, from
(\ref{inducadim2}) and for $\widetilde{\mu}=0$, the Fourier coefficients $b^{\pm}_n(r)$ must satisfy 
\begin{equation}
	[\gamma + k^2 + i(\overline{R}_m\overline{\mu}h(r)+ n \omega_f)] b^{\pm}_n
	 ={\cal L}^{\pm} b^{\pm}_n.
	\label{inducfourier}
\end{equation}
In addition,
the boundary condition (\ref{saut}) with  $f(t)=\cos(\tau)$ implies 
\begin{equation}
[D b_n^{\pm}]_{1+}^{1-} \pm \frac{i}{2}\overline{R}_m (b_n^+ + b_n^-)_{r=1}
\pm \frac{i}{4} \widetilde{R}_m (b_{n-1}^+ +b_{n-1}^- + b_{n+1}^+ +b_{n+1}^-)_{r=1} =0.
\label{saut3fourier}
\end{equation}

The solutions of (\ref{inducfourier}) which are continuous at $r=1$ can be written in the form
\begin{equation}
	b^{\pm}_n = C^{\pm}_n \psi^{\pm}_n, \;\; \mbox{with} \quad
	\psi^{\pm}_n=\left\{ \begin{array}{c}
\!\!\! I^{\pm}(q_nr)/ I^{\pm}(q_n), \quad r<1 \!\!\! \\
\!\!\! K^{\pm}(s_nr)/ K^{\pm}(s_n), \quad r>1 \!\!\! \\
\end{array} \right.,
\label{bpmfourier}
\end{equation}
with
\begin{equation}
q_n^2 = k^2 + \gamma +i(\overline{R}_m\overline{\mu}+n \omega_f), \quad s_n^2 = k^2 + \gamma + in \omega_f.
\label{qnsn}
\end{equation}

Substituting (\ref{bpmfourier}) in (\ref{saut3fourier}), we obtain the following system
\begin{equation}
	C_n^{\pm}{\cal R}_n^{\pm} \pm i \frac{\overline{R}_m}{2}(C_n^+ +C_n^-) \pm i \frac{\widetilde{R}_m}{4}  (C_{n-1}^+ +C_{n-1}^- + C_{n+1}^+ + C_{n+1}^-)) =0
	\label{systeme}
\end{equation}
with ${\cal R}^{\pm}_n= q_n I_n^{\pm '} / I_n^{\pm} - s_n K_n^{' \pm} / K_n^{\pm}$ 
and where $I_n^{\pm}=I_{m \pm 1}(q_n)$ and $K_n^{\pm}=K_{m \pm 1}(s_n)$
are modified Bessel functions of first and second kind.

The system (\ref{systeme}) implies the following matrix dispersion relation 
\begin{equation}
	{\cal F}_n C_n - i \frac{\widetilde{R}_m}{4} ({\cal R}^{+}_n - {\cal R}^{-}_n)  (C_{n-1} + C_{n+1}) =0
	\label{systeme2}
\end{equation}
with $C_j = C_j^+ + C_j^-$ and
\begin{equation}
	{\cal F}_n={\cal R}^+_n {\cal R}^-_n -i(\overline{R}_m/2)({\cal R}^+_n - {\cal R}^-_n).
	\label{Fn} 
\end{equation}
Solving the system (\ref{systeme2}) is equivalent to setting to zero the determinant of the matrix $A$ defined by
\begin{equation}
	A_{nn}={\cal F}_n \quad A_{n\;n-1}=A_{n\;n+1}= - i \frac{\widetilde{R}_m}{4} ({\cal R}^{+}_n - {\cal R}^{-}_n)
	\label{matA}
\end{equation}
and with all other coefficients being set to zero.
\subsection{High frequency regime for the periodic flow (i) with zero mean}
\label{highfrequency}
Following the notation of section \ref{resolution i},
and considering a periodic flow (i) with zero mean, we have
$\overline{\mu}=0$. From (\ref{qnsn}) this implies that $q_n=s_n$.
Using the identity
\begin{equation}
I_n^{\pm '} K_n^{\pm} - K_n^{' \pm} I_n^{\pm}=\frac{1}{s_n}
\end{equation}
we obtain ${\cal R}^{\pm}_n= (I_n^{\pm}K_n^{\pm})^{-1}$. 
As $\overline{R}_m=0$, the equation (\ref{Fn}) becomes ${\cal F}_n={\cal R}^+_n {\cal R}^-_n$. 
Then we can rewrite the system (\ref{systeme2}) in the form
\begin{equation}
       C_n + i \frac{\widetilde{R}_m}{4} (I_n^{+}K_n^{+} - I_n^{-}K_n^{-}) (C_{n-1} + C_{n+1}) =0.
	\label{systeme0}
\end{equation}
From the asymptotic behaviour of the Bessel functions for high arguments, we have
\begin{equation}
\alpha_n \equiv I_n^{-}K_n^{-} - I_n^{+}K_n^{+} \approx 1/s_n^3.
\end{equation}
For the high values of $n$ these terms are negligible and in first approximation
we keep in the system (\ref{systeme0}) only the terms corresponding to $n=0, \pm1$. 
This leads to a $3 \times 3$ matrix system whose determinant is
\begin{equation}
1 + \frac{(\widetilde{R}_m)^2}{16}\alpha_0(\alpha_{-1} + \alpha_1)=0.
\label{syst33}
\end{equation}
At high forcing frequencies $\omega_f$ we have $s_{\pm1} \approx \sqrt{\omega_f}$. Together with (\ref{syst33}), it implies 
\begin{equation}
	\widetilde{R}_m \approx \omega_f^{3/4}.
\end{equation}
In addition, from the approximate $3 \times 3$ matrix system, the double-frequency $2 \omega_f$
emerges for $n=\pm 1$.

\subsection{High frequency regime and small modulation amplitude for the periodic flow (i) with non zero mean}
\label{highfrequency2}
For small amplitude modulation $\rho<<1$ the system 
(\ref{systeme}) is truncated so as to keep the first Fourier modes $n=0$
and $n=\pm 1$. The dispersion relation :
\begin{equation}
{\cal F}_0+ \rho^2 \left({\overline{R}_m \over 4}\right)^2 ({\cal R}^{+}_{0}-{\cal R}^{-}_{0})
\left({{\cal R}^{+}_{-1}-{\cal R}^{-}_{-1}
\over {\cal F}_{-1}}+{{\cal R}^{+}_{+1}-{\cal R}^{-}_{+1} \over
{\cal F}_{+1}}\right)=0 \label{disp}
\end{equation}
is then solved perturbatively setting $\overline{R}_m=R_0 +\delta R$ and
$\overline{\omega}=\omega_0 +\delta \omega$ and expanding ${\cal F}_0(\overline{R}_m,
\overline{\omega})$
to first order in $\delta R$ and $\delta \omega$ given that 
${\cal F}_0(R_0,\omega_0)=0$ and with the constants $C_0^{\mp}=\pm {\cal R}^{\pm}_{0}$.
The  dispersion  relation (\ref{disp}) becomes
\begin{equation}
\delta R {\partial {\cal F}_0 \over \partial \overline{R}_m} + \delta \omega 
{\partial {\cal F}_0 \over \partial \overline{\omega}}=
-\rho^2 \left({R_0 \over 4}\right)^2 C_0\left({\beta_{-1}\over {\cal F}_{-1}}+
{\beta_{+1}\over {\cal F}_{+1}}\right) \label{R2}
\end{equation}
with  $\beta_n={\cal R}^{+}_n - {\cal R}^{-}_n$. The threshold and frequency 
shifts which behave like $\rho^2$ are written $\delta R=\rho^2 R_2$ 
and $\delta \omega=\rho^2 \omega_2$. In the left-hand-side of (\ref{R2}) the partial 
derivatives are given by
\begin{eqnarray}
{\partial {\cal F}_0 \over \partial \overline{\omega}}&=&-{1 \over C_0}\left[(C_0^+)^2
{\partial {\cal R}_0^+ \over \partial \overline{\omega}}- (C_0^-)^2
{\partial {\cal R}_0^- \over \partial \overline{\omega}}\right] \\
{\partial {\cal F}_0 \over \partial \overline{R}_m}&=&-{1 \over C_0}\left[(C_0^+)^2
{\partial {\cal R}_0^+ \over \partial \overline{R}_m}-(C_0^-)^2
{\partial {\cal R}_0^- \over \partial \overline{R}_m}\right]-{i \over 2} C_0  
\end{eqnarray}
One can show that the partial derivatives of ${\cal R}_0^{\pm}$ are related to integrals  
calculated in \cite{Normand03} through the relations 
\begin{equation}
{\partial {\cal R}_0^{\pm} \over \partial \omega}\equiv i \int_0^{\infty} (\Psi_0^{\pm})^2 r dr
\quad \quad
{\partial {\cal R}_0^{\pm} \over \partial R_m} \equiv i \overline{\mu} \int_0^1 (\Psi_0^{\pm})^2 r dr
\label{pder}
\end{equation}
In the following we shall focus on the case $\overline{\mu}=0$ and we introduce
the notations
\begin{equation}
{\partial {\cal F}_0 \over \partial \overline{R}_m}=-i C_0 (f_1 +i f_2), \quad 
{\partial {\cal F}_0 \over \partial \overline{\omega}}= -i C_0 (g_1 +i g_2),\quad
{\beta_{-1}\over {\cal F}_{-1}}+
{\beta_{+1}\over {\cal F}_{+1}}=X+iY
\end{equation}
Solutions of (\ref{R2}) are
\begin{equation}
 R_2= \left({R_0 \over 4}\right)^2{X g_1 + Y g_2 \over f_1 g_2 -f_2 g_1} \quad \quad
 \omega_2 = \left({R_0 \over 4}\right)^2{X f_1 + Y f_2 \over f_1 g_2 -f_2 g_1} \label{shift}
\end{equation}
recovering results similar to those obtained in \cite{Normand03} using a different approach. 

We have in mind that for some values of $\omega_f$ resonance can occur.
An oscillating system forced at a resonant frequency is prone to instability 
and a large negative threshold shift is expected.
However, inspection of (\ref{shift}) reveals no clear relation between
the sign of $R_2$ and the forcing frequency which appears in the 
quantities $X$ and $Y$. We only know that $f_1 g_2 -f_2 g_1<0$,
since near the critical point ($\delta R=\overline{R}_m-R_0$) the 
denominator in (\ref{shift}) is proportional to the growth rate
of the dynamo driven by a steady flow. When $\rho=0$,
 we shall consider Eq. (\ref{R2}) for an imposed $\delta R$ and complex 
values of $\delta \omega=\omega_1 + i \sigma_1$ where $\omega_1$ is the 
frequency shift and $\Re(\gamma)=-\sigma_1$ is the growth rate, given by
\begin{equation}
\sigma_1=\delta R {f_1 g_2 -f_2 g_1 \over g_1^2 + g_2^2}
\end{equation}
Above the dynamo threshold, ($\delta R>0$) the field is amplified
($\Re(\gamma)>0$) thus $\sigma_1<0$ and $f_1 g_2 -f_2 g_1<0$.

In the high frequency limit ($\omega_f>>\omega_0$) expressions for $X$ and $Y$ 
can be derived explicitly using the asymptotic behaviour of the
Bessel functions for large arguments. For $\overline{\mu}=0$
with $q_{\pm 1}=s_{\pm 1}\approx(\omega_f \pm \omega_0)^{1/2}(1 \pm i)/\sqrt 2$ 
and using the asymptotic behaviour : $\beta_{\pm 1}/ {\cal F}_{\pm 1} \to (s_{\pm 1})^{-3}$
one gets
\begin{equation}
X+i Y
=-{\sqrt 2} \omega_f^{-3/2}(1 - i{3\omega_0 \over 2 \omega_f}) \label{HF1}
\end{equation}
When $\overline{\mu}=0$, we have also $f_1=1/2$ and $f_2=0$, leading to the expression for $R_2$ :
\begin{equation}
R_2 \approx -{R_0^2 \over 4 {\sqrt 2}} \omega_f^{-3/2} 
\left({g_1 \over g_2}-{3\omega_0 \over 2 \omega_f} \right)
\end{equation}
For the wave numbers $m=-k=1$, numerical calculations of $g_1$ and $g_2$ which only 
depend on the critical parameters $R_0$ and $\omega_0$ give $g_1/g_2=1.626$, 
and thus $R_2<0$ when $\omega_f \to \infty$. 

When $\overline{\mu}=0$, there are several reasons to consider the particular value of 
the forcing : $\omega_f=2\omega_0$. One of them is that for Hill or Mathieu 
equations it is a resonant frequency. Moreover, in the present
 problem it leads to simplified calculations. In particular 
the asymptotic behavior of $\beta_n/{\cal F}_n$ can still be used for $n=+1$
since $\omega_f+\omega_0$ is large, while the approximation is no longer valid
for $n=-1$. Nevertheless, the mode $n=-1$ is remarkable since it corresponds 
to $s_{-1}=s_0^*$ from which it follows that
$\beta_{-1}=\beta_0^*$ and ${\cal F}_{-1}=-i R_0 \beta_0^*$.
Finally one gets the exact result : $\beta_{-1}/ {\cal F}_{-1}=i/R_0$,
which leads to
\begin{equation}
X+i Y \approx {i \over R_0}+ {1\over s_1^3}\quad \mbox{with} \quad 
s_1^3\approx -2\left(3\omega_0\over 2\right)^{3/2}(1-i)
\end{equation}
For the values $R_0=20.82$ and $\omega_0=4.35$ corresponding to Fig. \ref{fig:R2km1} (f) one gets
$X=-1.5 \times 10^{-2}$ and $Y=3.3\times 10^{-2}$. The threshold shift is
\begin{equation}
R_2 \approx {R_0^2 \over 8}(1.62X+Y)=0.48
\end{equation}
showing that the sign of $R_2$ changes when $\omega_f$ decreases from infinity to
$2\omega_0$. This result is in qualitative agreement with the exact 
results reported in Fig. \ref{fig:R2km1} (f) where
$R_2=0$ for $\omega_f=8.3\simeq 2\omega_0$. When the forcing frequency is 
exactly twice the eigen-frequency $\omega_0$ we  had rather expected 
a large negative value of $R_2$ on the basis it is a resonant condition
for ordinary differential system under temporal modulation. 
In Fig. \ref{fig:R2km1} (f) the maximum negative value of $R_2$ occurs for 
$\omega_f\simeq 4 \omega_0$ which cannot be explained by simple arguments.

For $\overline{\mu}\ne 0$, we have not been able to find resonant conditions
like : $n \omega_f + m \omega_0=0$ ($n$, $m$ integers) between $\omega_f$ 
and $\omega_0$ such that $\omega_f$ would be
associated to a special behavior of the threshold shift.
 Contrary to the Hill equation, the induction
equation is a partial differential equation with the consequence that the  
spatial and temporal properties of the dynamo are not independant. The 
wave numbers $k$ and $m$ are linked to the frequencies $\omega_0$ and
$\omega_f$
through $q_{\pm n}$ and $s_{\pm n}$ which appears as arguments of 
Bessel functions having rules of composition less trivial than
trigonometric functions. Exhibiting resonant conditions
implies to find relationship between $q_{\pm n}$, $s_{\pm n}$ and $q_0$, $s_0$ 
for specific values of $\omega_f$. We have shown above for $\overline{\mu}=0$
that a relation of complex conjugaison exists for $n=-1$ when $\omega_f=2\omega_0$
but we have not yet found how to generalize to other values of $\overline{\mu}$
and have left this part for a future work.

\subsection{Resolution of case (ii): smooth flow}
\label{resolution ii}
We define the trial functions $\psi^{\pm}_j = K_{m\pm1}(k) J_{m\pm1}(\alpha_j r) / J_{m\pm1}(\alpha_j)$ 
  where the $\alpha_j$ are the roots of the equation  
\begin{equation}
	\alpha_j \left[ \frac{K_{m+1}(k)}{J_{m+1}(\alpha_j)} - \frac{K_{m-1}(k)}{J_{m-1}(\alpha_j)} \right] + 2k\frac{K_m(k)}{J_m(\alpha_j)} = 0,
	\label{alphaj}
\end{equation}
and where $J$ and $K$ are respectively the Bessel functions of first kind and the modified Bessel functions of second kind.
For $r \le 1$, we look for solutions in the form 
\begin{equation}
	b^{\pm} = \sum_{j=1}^{N} b^j(t) \psi^{\pm}_j (\alpha_j r)
	\label{troncature}
\end{equation}
where $N$ defines the degree of truncature.
For $r \ge 1$ the solutions of (\ref{domextii}) are thus of the form 
\begin{equation}
	b^{\pm} = K_{m\pm 1} (k r) \sum_{j=1}^{N} b^j(t) 
\end{equation}
and, from (\ref{alphaj}) these solutions satisfy the conditions (\ref{saut1}) at the interface $r=1$.
To determine the functions $b_j(t)$ it is sufficient to solve the induction equation (\ref{inducadim})
for $r \le 1$. 
For that we replace the expression of $b^{\pm}$ given by (\ref{troncature}) into the induction equation
(\ref{inducadim}), in order to determine the residual
\begin{equation}
	R^{\pm} = \sum_{j=1}^{N} \left(\dot b_j + [k^2 + \alpha_j^2
+ i (m \Omega + k V) ]b_j    \right)\psi^{\pm}_j 
 \mp  \frac{i}{2}r \Omega' \sum_{j=1}^{N}   b_j(\psi^{+}_j  + \psi^{-}_j ).
\end{equation}
Then we solve the following system
\begin{equation}
	\int_0^1 R^+ \phi^{+}_i r dr + \int_0^1 R^- \phi^{-}_i r dr = 0, \quad i=1, \cdots, N
	\label{syst}
	\end{equation}
where the weighting functions are defined by $\phi^{\pm}_j = J_{m\pm1}(\alpha_j r) / J_{m\pm1}(\alpha_j)$.
Using the orthogonality relation
\begin{equation}
	\int_0^1 \phi^{+}_i \psi^{+}_j r dr + \int_0^1 \phi^{-}_i \psi^{-}_j r dr = \delta_{ij} G_{ij}
\end{equation}
with
\begin{equation}
	G_{ii} = \frac{K_{m+1}(k)}{J_{m+1}^2 (\alpha_i)} \int_0^1 J_{m+1}^2 (\alpha_i r) r dr + 
	\frac{K_{m-1}(k)}{J_{m-1}^2 (\alpha_i)} \int_0^1 J_{m-1}^2 (\alpha_i r) r dr,
\end{equation}
we write the system (\ref{syst}) in the following matrix form
\begin{equation}
	  \dot{\textbf{X}} = M\textbf{X} \quad \mbox{with} \quad \textbf{X}=(b_1, \cdots, b_N)
\end{equation}
with
\begin{eqnarray}
	\quad M_{ij}=\delta_{ij} (k^2 + \alpha_j^2) &+& \frac{i}{G_{ii}}\int_0^1 (\phi^{+}_i \psi^{+}_j + \phi^{-}_i \psi^{-}_j)(m\Omega + k V)r dr \\
	&-& \frac{i}{2 G_{ii}}\int_0^1 (\phi^{+}_i - \phi^{-}_i)(\psi^{+}_j +  \psi^{-}_j)\Omega' r^2dr.
\end{eqnarray}
The numerical resolution of this system is done with a fourth order Runge-Kutta time-step scheme. We took a white noise  as an initial condition for the $b_j$.
\subsection{Corrigendum of Normand (2003) results \cite{Normand03}}
\label{Corrigendum}
For very small values of the fluctuation rate $\rho$ and for an infinite shear (case (i)), a
comparison can be made between the results obtained for $\widetilde \mu=0$ by 
the method based on Floquet theory (see Appendix \ref{resolution i}) and those  
obtained by a perturbative approach  \cite{Normand03} which consists 
in expanding
$\overline{R}_m$ and the frequency $\Im(p)$ in powers of $\rho$ according to
\begin{eqnarray}
\Im(p) &=& \omega_0  + \rho \omega_1 + \rho^2 \omega_2 + ....\\
\overline{R}_m &=& R_0 + \rho R_1 + \rho^2 R_2 + ....
\end{eqnarray}
where $R_0$ and $\omega_0$ are respectively the critical values of 
the Reynolds number and the frequency in the case of a stationary flow.
At the leading order it appears that : $\omega_1=R_1=0$. At the next order, the  
expressions of $R_2$ and $\omega_2$ are given in \cite{Normand03}, however 
their numerical values are not correct due to an error 
 in their computation.\\ 
\begin{figure}
\begin{tabular}{@{\hspace{-2cm}}l}
    \includegraphics[width=1.15\textwidth]{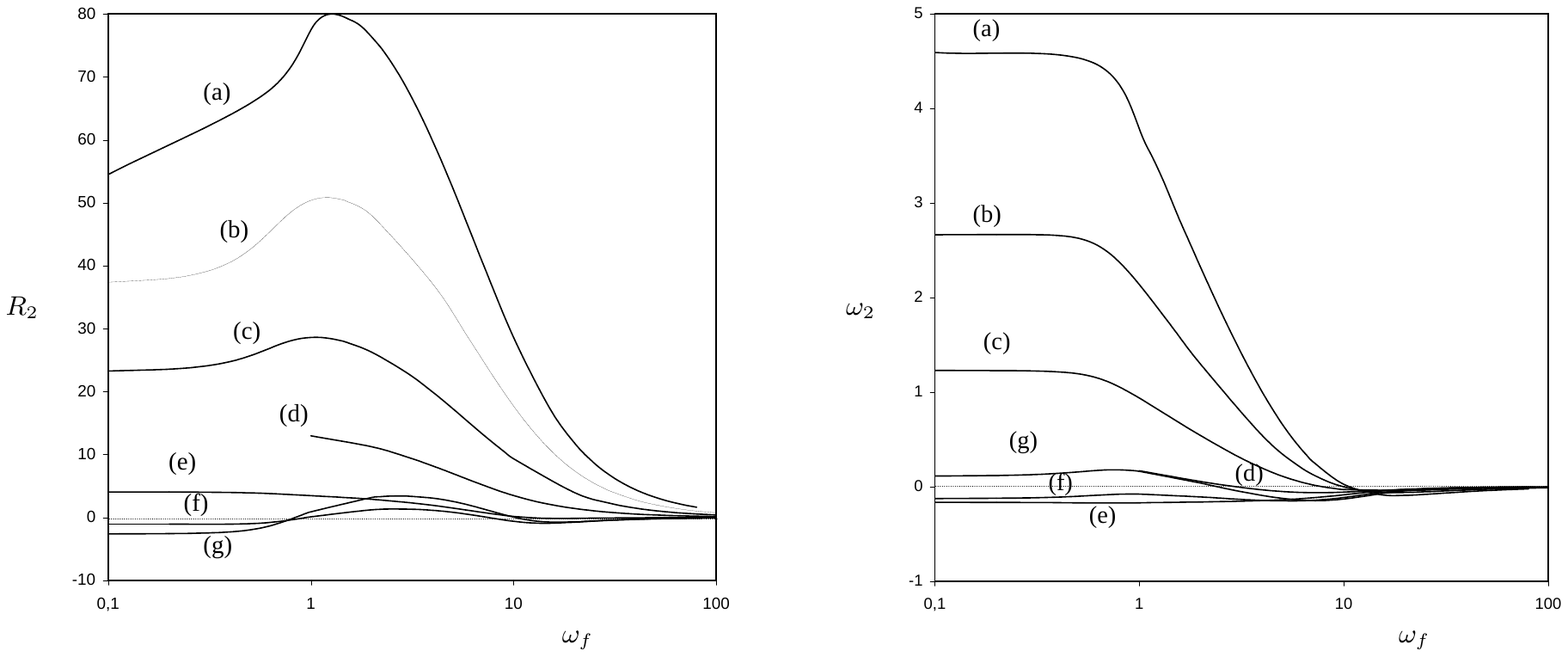}\\*[-20cm]
\end{tabular}  
    \caption{Results obtained by the perturbative approach for $m=1$, $k=-0.56$ and $\overline \mu=0.44$ ($\overline \Gamma=1$). The dynamo threshold $R_2$ is plotted versus $\omega_f$ for several values of $\widetilde{\mu} = $
(a) 1.56; (b) 1.28; (c) 1; (d) 0.72; (e) 0.44; (f) 0.16 and (g) 0.}
\label{fig:R2km0p56}
\end{figure}
After correction, the new values of $R_2$ and $\omega_2$ are given in Figure \ref{fig:R2km0p56} for the set of parameters considered in \cite{Normand03}:
 $m=1$, $k=-0.56$ and $\overline \Gamma=1$ ($\overline \mu=0.44$). 
 The different curves correspond to values of $\widetilde \Gamma$
 which are not necessarily the same than those taken in \cite{Normand03}. 
For $|\widetilde{\mu}|$ sufficiently small (curves e and f), $R_2$ changes its sign 
twice versus the forcing frequency.
We find that $R_2$ is negative for low and high frequencies implying a dynamo threshold smaller than the one for the stationary flow. At intermediate frequencies $R_2$ is positive with a maximum value, implying a dynamo threshold larger than the one for the stationary flow.
For larger values of $|\widetilde{\mu}|$ (curves a, b, c and d)
$R_2$ is positive for all forcing frequencies
with a maximum value at a low frequency which increases with
$|\widetilde \mu|$. 
This implies that for $|\widetilde \mu|$ sufficiently large the dynamo threshold is larger than the one obtained for the stationary flow as was already mentioned
in section \ref{section:stationary}.
For $\widetilde \Gamma=1.78$ we have
$\widetilde \mu=0$. In this case, we have 
checked that the values of $R_2$ and $\omega_2$ are in good agreement with the  
values of $R_m$ and $\overline \omega$ obtained by the method of Appendix \ref{resolution i}, 
provided $\rho \le 0.1$.
\begin{figure}
\begin{tabular}{@{\hspace{-2cm}}l}
    \includegraphics[width=1.15\textwidth]{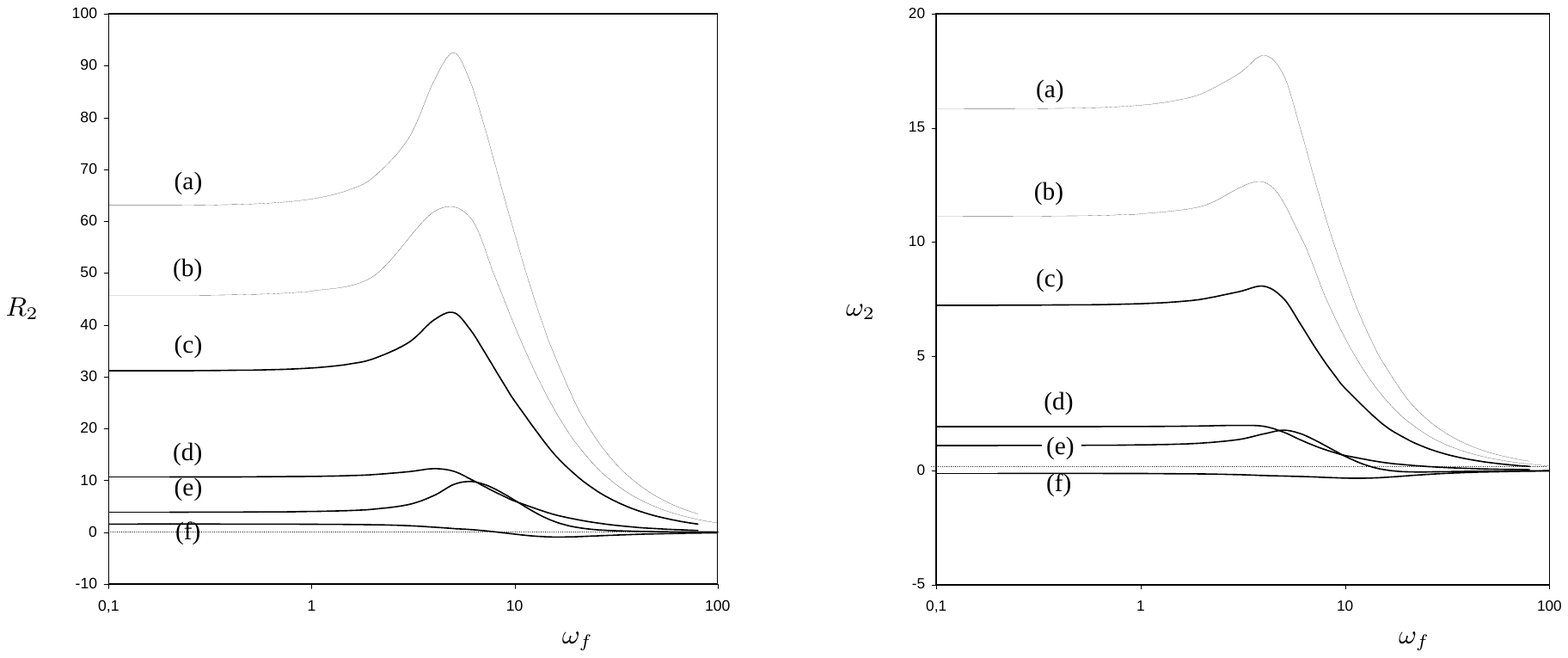}\\*[-20cm]
\end{tabular}  
\caption{Same as Figure \ref{fig:R2km0p56} but for $m=1$, $k=-1$ and $\overline \mu=0$ ($\overline \Gamma=1$).
The labels correspond to $\widetilde{\mu} = $ (a) 1.5; (b) 1.25; (c) 1; (d) 0.5; (e) -0.5 and (f) 0.}
\label{fig:R2km1}
\end{figure}

For completeness we have also calculated the values of $R_2$ and $\omega_2$ for
$m=1$, $k=-1$ and $\overline \Gamma=1$, ($\overline \mu=0$) as considered 
in the body of the paper. The results are plotted in Figure \ref{fig:R2km1}.
Qualitatively the results are in good agreement with those of  Figure \ref{fig:R2km0p56}.
 For $\widetilde \mu=0$ again, we have 
checked that the values of $R_2$ and $\omega_2$ are in good agreement with the  
values of $R_m$ and $\overline \omega$ obtained by the method of Appendix \ref{resolution i}, 
provided $\rho \le 0.1$. For higher values of the modulation amplitude $\rho$
the relative difference between the results obtained by the two methods can reach 10\%
on $R_2$ for $\rho=0.4$.

Finally it must be noticed that our parameters $\widetilde \Gamma$ and $\omega_f$
are strictly equivalent to respectively $\varepsilon_1$ and $\sigma$
in \cite{Normand03}.

\end{document}